\documentclass[aps,pra,preprint,groupedaddress,showpacs]{revtex4}

\usepackage{graphicx}
\usepackage{amsmath}
\usepackage{amsfonts,amssymb}
\newcommand{\ket}[1]{| #1 \rangle}
\newcommand{\bra}[1]{\langle #1 |}
\newcommand{\rb}[1]{\left( #1 \right)}

\begin{document}


\title {Time evolution of the Rabi Hamiltonian from the
unexcited vacuum}


\author{R. F. Bishop and C. Emary}
\email[]{emary@theory.phy.umist.ac.uk}
\affiliation{ Department of Physics,
           UMIST,
           P.O. Box 88,
           Manchester 
           M60 1QD, 
	   U. K.}


\date{\today}

\begin{abstract}
The Rabi Hamiltonian describes a single mode of electromagnetic
radiation interacting with a two-level atom.  Using the coupled
cluster method, we investigate the time evolution of this system
from an initially empty field mode and an unexcited atom.  We give
results for the atomic inversion and field occupation, and find
that the virtual processes cause the field to be squeezed. No
anti-bunching occurs.
\end{abstract}

\pacs{03.65.Ca, 31.70.Hq, 32.80.-t, 42.50.Dv}

\maketitle

\section{Introduction}

\normalsize

The Rabi Hamiltonian plays an important role in quantum optics. It
describes a two-level atom interacting with a single mode of
quantized electromagnetic radiation via a dipole interaction
\cite{al:eb}. It also finds wider application, describing a spin
interacting with phonons in NMR \cite{ii:ra}, for example.  It is
also related to the static Lee model in field theory \cite{ma:pe}.

The field mode is described by bosonic annihilation and creation
operators, $b$ and $b^\dagger$ respectively, which obey the usual commutation relation,
\begin{equation}
\left[b,b^{\dagger}\right] = 1.
\end{equation}
The two-level atom is described by the pseudo-spin operators
\begin{equation}
\sigma^z=\rb{\begin{array}{lr}1 & 0\\0 & -1 \end{array} } ,~~
\sigma^+=\rb{\begin{array}{lr}0 & 2\\0 & 0 \end{array} } ,~~
\sigma^-=\rb{\begin{array}{lr}0 & 0\\2 & 0 \end{array} };
\end{equation}
\begin{equation}
\sigma^x = \frac{1}{2}\left(\sigma^+ + \sigma^-\right),~~
\sigma^y = \frac{i}{2}\left(\sigma^- - \sigma^+\right).
\end{equation}
With these definitions the Rabi Hamiltonian is given by
\begin{equation}
H = \frac{1}{2} \omega_{0} \sigma^z + \omega b^{\dagger}b + g \left(\sigma^{+} + \sigma^{-}\right)
       \left(b^{\dagger} + b \right).           \label{Hamiltonian}
\end{equation}
There is a conserved parity $\Pi$ associated with the Hamiltonian,
\begin{eqnarray}
\Pi \equiv \exp \rb{ i \pi N }, & N \equiv
b^\dagger b + \frac{1}{2}\left(\sigma^z + 1\right). \label{parity}
\end{eqnarray}

There is no proof that this Hamiltonian is integrable, although
suggestive evidence does exist \cite{re:la}. Consequently,
investigation of the Rabi system requires approximations to be
made.

The most widely used approach is to make the rotating wave
approximation (RWA).  This was first applied by Jaynes and
Cummings \cite{ja:cu}, yielding the Jaynes--Cummings Model. Under
the RWA, one neglects the ``virtual terms''
$\sigma^{+}b^{\dagger}$ and $\sigma^{-}b$, also known as the
counter-rotating terms.  This leads to the excitation number $N$
as well as parity $\Pi$ being conserved, and renders the model
soluble by a series of $2\times 2$ diagonalisations.

Despite its frequent use in quantum optics the
actual validity of the RWA for specific applications is usually
highly questionable.  For example, we know \cite{fe:ra} that the energy
spectrum of the Rabi Hamiltonian can be approximated by its JC counterpart in
the RWA only for sufficiently small values of the coupling
strength $g$, and that the width of this range decreases as one
proceeds higher up the spectrum.  There is a further 
concern regarding the RWA.  Ford and O'Connell
\cite{fo:oc} investigated the system of a charged harmonic
oscillator interacting with a reservoir consisting of an infinite
number of oscillators within the RWA.  They showed that the
spectrum has no lower bound for all models of physical interest.
This constitutes a violation of the second law of thermodynamics,
as we can take energy from the oscillator bath (modeling the
environment) without producing an effect upon it, i.e., we can
still remove an infinite amount of energy from it.

The energy level spectrum of the full Hamiltonian has been
investigated by several authors.  The simplest approach is to use
the configuration-interaction (CI) method, equivalent to a
large-scale diagonalisation in a suitably defined finite subspace
of the full Hilbert space. This method has been used both by Graham and
H\"{o}hnerbach \cite{gr:ho} and by Ku\'{s} \cite{ku:sc} in
investigating possible quantum chaos signatures of the model. Lo
{\it et al.} \cite{lo:li} have given an analysis of the validity of the
CI method. Reik and others \cite{re:nu} have adapted Judd's method
\cite{ju:dd} for the Jahn-Teller system for use with the Rabi Hamiltonian.
Here, the Hamiltonian is translated into the
Bargmann representation \cite{ba:rg} and solutions of the
resulting differential equations are sought. Whereas Judd
originally used a power series Ansatz, Reik uses a Neumann series.
For certain couplings, the Neumann series terminates, giving
isolated, exact solutions known as Juddian solutions. Elsewhere
the series gives a useful, convergent approximation. The Juddian
points are valuable for comparison of approximate techniques.
Variational results have also been provided by Bishop {\it et al.}
\cite{bi:va} and by Benivegna and Messina \cite{be:me}.  The
latter method also permits perturbative corrections, allowing the
exact results to be approached.

The time evolution of the Rabi Hamiltonian has been of considerable interest
for a long time.  We mention in particular the pioneering studies of Shirley
\cite{jh:sh}. Somewhat later, Eberly and co-workers \cite{eb:na} were the
first to fully demonstrate the rich time evolution of this model. They worked
within the RWA and discovered the collapse-revival nature of the subsequent
evolution when the system is started from a coherent field \cite{mi:si}.  The
impact of the non-RWA terms on the time evolution has been investigated using
various techniques. For example, Zaheer and Zubairy have used path-integral
methods \cite{za:zu}, while several authors have used perturbative techniques
\cite{fa:zh,ph:nx}.  The drawback of these latter methods is their restriction
to small coupling. Finally, Swain has given formally exact result for the time
evolution of the Rabi Hamiltonian \cite{sw:jp}.  However, the solutions are
expressed as continued fractions, which limits their usefulness in practice.

In this paper we shall investigate the evolution of the Rabi
system from an initial state composed of an empty field mode and
an unexcited atom, which we denote as $\ket{0, \downarrow}$ and henceforth
call the unexcited vacuum state.  Within the RWA, the system would
remain in this state indefinitely. However, inclusion of the
counter-rotating terms means that we allow ``energy
non-conserving'' processes to occur.  It is these virtual
processes that drive the evolution of the system. For simplicity,
we shall only consider the resonant case, $\omega=\omega_0$. 

In order to investigate the evolution of the Rabi system from the state
$\ket{0,\downarrow}$ we employ here one of the most versatile and most
accurate semi-analytical formalisms of microscopic quantum many-body theory,
namely the coupled cluster method (CCM)
[23--30].  Coupled cluster
techniques are widely regarded as being amongst the most powerful of all {\it
ab initio} quantum many-body methods.  As such they provide a number of
distinct advantages over more traditional or more specialised methods which
have hitherto been used in quantum optics and allied fields and, more
specifically, to study Rabi systems.  The CCM exists in two versions, the
so-called normal (NCCM) and extended (ECCM) types \cite{ja:rp}, and in the
present paper we employ only the former version.

A particular advantage of the CCM, a nonperturbative method originally
developed in nuclear physics by Coester and K\"{u}mmel \cite{co:ku}, is that
it has been extremely widely applied and tested on a huge variety of physical
systems \cite{bi:tc}.  These include areas as diverse as nuclear and
subnuclear physics, quantum field theories (both in the spatial continuum and
on the spatial lattice), condensed matter physics, quantum magnetism, and
quantum chemistry.  In almost all such cases CCM techniques now provide
numerical results that are either the best or among the best available.

A pertinent, but quite typical, example is the electron gas,
one of the most intensely studied of all quantum many-body problems.  Here,
the CCM results \cite{bi:lu, bi:lp, em:za} for the correlation energy, for
example, agree over the entire metallic density range to within less than one
millihartree per electron (or $<1\%$) with the essentially exact Greens's
function Monte Carlo results that are, very exceptionally, available for this
fermionic system. The CCM results have never been bettered by any other
technique. Elsewhere, in quantum chemistry, for example, after its early
introduction by \v{C}i\v{z}ek \cite{ci:ze}, the influence of the CCM has been
profound, to the point where it it nowadays the method of choice for most
highly accurate chemical studies \cite{ja:po, rj:ba}.

Since the Rabi Hamiltonian involves a two-level system which is modeled by
Pauli pseudospin operators, we note finally that the CCM has also been very
successfully applied to a large number of spin-lattice systems exhibiting
anti-ferromagnetic and other forms of magnetic ordering.  In instances such as
unfrustrated models on bipartite lattices where a Marshall-Peierls sign rule
\cite{wm:ar} exists (which provides a means to circumvent the infamous
``minus-sign problem'' inherent in simulating many-fermion systems by
quantum Monte Carlo (QMC) techniques), the CCM provided results fully
comparable in accuracy with those obtained by QMC means \cite{cz:fb}. The CCM
is even able to predict with good accuracy the positions of the phase
boundaries which mark the quantum phase transitions between states of
different quantum order. Furthermore, the CCM provides equally accurate
results for systems which are frustrated either geometrically (e.g., on a
triangular lattice) \cite{cz:fb}, or dynaimically (e.g., by competing
interactions on different bonds on the lattice) \cite{kr:sf}, where QMC
methods are much more difficult to apply. In several such instances the CCM
results are now the best available.

Although the time-dependent formalism of the CCM \cite{ja:rp} has existed for
a considerable time, it has not yet found the same range of applications as
its static counterpart. For example, Hoodbhoy and Negele \cite{ho:ne} have
investigated the application of the technique in nuclear dynamics, using two
interacting Lipkin Hamiltonian systems as a test model.  It has also found
more application in chemical physics \cite{sg:dm, qc:mo}.  For example,
Monkhorst has outlined the application of the time-dependent CCM to the
treatment of molecular eigenstates \cite{hj:mo}, with the aim to describe such
phenomena as scattering, chemical dynamics, and laser chemistry.  Sree Latha
and Durga Prasad have investigated the application of the technique to
multi-mode systems with vibronic coupling \cite{sl:dp}, a mechanism
describing, amongst other things, non-adiabatic phenomena in the electronic
spectroscopy of polyatomic molecules.  We note that all of these studies have
tended to only use the lowest levels of CCM approximation, due to the
complexity of the systems studied.  In comparison, we shall use high levels of
approximation in an endevour to describe the Rabi system as accurately as
possible.

For other applications of the CCM the interested reader is referred
to the reviews contained in Refs. \cite{bi:tc, bi:vl}.  An important feature
of all the applications is that the results provided by the CCM are able to be
systematically improved upon via well-defined hierachies of approximations.
At each level of approximation the results are open to ready physical
interpretation in terms of the correlated many-body clusters involved and
their corresponding multiconfigurational creation operators.  To date the NCCM
has already been used to investigate the ground-state of the Rabi Hamiltonian
\cite{bi:cc}, where it has been shown to give excellent results for values of
the coupling parameter $g\lesssim 0.665$.
In view of the demonstrable success in treating so many other many-body
systems, we now wish to apply it to the dynamical evolution of Rabi systems,
with a particular aim to shed light on the importance of the counter-rotating
terms dropped in the commonly made RWA.

\section{Formalism}
\normalsize 
We now
briefly describe the application of the normal coupled cluster method 
to the Rabi Hamiltonian of interest to us here.

Let $\ket{\Psi(t)}$ and $\langle\widetilde{\Psi}(t)|$ be the exact
ket and bra states at time $t$ for an arbitrary many-body quantum
mechanical system, chosen so that
\begin{equation}
\langle\widetilde{\Psi}(t)\ket{\Psi(t)} = 1, ~~\forall~t.
\label{norm}
\end{equation}
The Hilbert space for our system may be described in terms of a
model state or cyclic vector  $\ket{\Phi_0}$ and a corresponding
complete set of mutually commuting multiconfigurational creation
operators $\left\{C_I^{\dagger}\right\}$.  The set
$\left\{C_I^{\dagger}\right\}$ is defined with respect to the
reference state, such that
$C_I\ket{\Phi_0}=0=\langle\Phi_0|C_I^{\dagger},~~\forall ~I\ne 0$, in
a notation in which $C_0^{\dagger} \equiv 1$, the identity
operator. In general, $I$ is a set index and the operators
$C_I^{\dagger}$ contain products of single-particle operators. The
set index $\left\{ I \right\}$ is complete in the sense that the
set of states $\left\{C^{\dagger}_I\ket{\Phi_0}\right\}$ provides
a complete basis for the Hilbert space.  The reference state,
$\ket{\Phi_0}$, must be chosen to be non-orthogonal to the actual
wavefunction of the system,
\begin{equation}
\langle \Phi_0 \ket{\Psi(t)}\ne 0, ~~\forall~t.         \label{PhiPsi}
\end{equation}

Usually one chooses $\langle \Phi_0 | C_I C^{\dagger}_J
\ket{\Phi_0}=\delta_{IJ}$, where $\delta_{IJ}$ is some suitably
defined Kronecker symbol. However, our later choice of
$\left\{C^{\dagger}_I \right\}$  will leave the set
$\left\{C^{\dagger}_I\ket{\Phi_0}\right\}$ orthogonal but not
normalised;
\begin{equation}
\langle \Phi_0 | C_I C^{\dagger}_J \ket{\Phi_0}=N_I\delta_{IJ}.
\label{Cnorm}
\end{equation}

The ket and bra states are formally parametrized independently in the normal
coupled cluster method as
\begin{eqnarray}
\ket{\Psi (t)} = e^{k(t)}e^{\hat{S}(t)}\ket{\Phi_0} , ~~
\langle\widetilde{\Psi}(t)| =
e^{-k(t)}\bra{\Phi_0}\hat{\widetilde{S}}e^{-\hat{S}},
\label{NCCMparam}
\end{eqnarray}
where
\begin{eqnarray}
\hat{S} = \sum_{I \ne 0} s_I(t) C_I^{\dagger} , ~~
\hat{\widetilde{S}} = 1+\sum_{I \ne 0} \widetilde{s}_I(t) C_I ,
\label{clustop}
\end{eqnarray}
and $k(t)$ is a {\it c}-number.
Using this parametrization, the expectation value of an arbitrary
operator, $\hat{X}$, is given by
\begin{equation}
 \langle \hat{X} \rangle = \langle\widetilde{\Psi }(t)|\hat{X}\ket{\Psi (t)}
 = \langle \Phi_0
|\hat{\widetilde{S}}e^{-\hat{S}}\hat{X}e^{\hat{S}}\ket{\Phi_0}. \end{equation}
In the evaluation of $\langle \hat{X} \rangle$, we use the nested commutator
relation, \begin{equation}
\tilde{\hat{X}} \equiv e^{-\hat{S}}\hat{X}e^{\hat{S}} = \hat{X} +
\left[\hat{X},\hat{S}\right] +
\frac{1}{2!}\left[\left[\hat{X},\hat{S}\right]\hat{S}\right] +
\cdots~. \label{BCHreln}
\end{equation}
This equation describes a similarity transformation of the operator $\hat{X}$.
The similarity transformed Hamiltonian, 
$\tilde{H} \equiv e^{-\hat{S}}He^{\hat{S}}$, lies at the heart of the coupled 
cluster method.The ability of the NCCM to describe the Rabi stationary ground
state has been investigated elsewhere \cite{bi:cc,wa:ph}.  Here we
shall only consider the time-dependent formalism.  To this end, we
introduce the action functional
\begin{equation}
A\equiv\int_{t_0}^{t_1}\left\{\langle\widetilde{\Psi}(t)|\left(i\frac{\partial
} { \partial t} - H \right)\ket{\Psi(t)}\right\}. \label{action}
\end{equation}
The stationarity principle,
\begin{equation}
\frac{\delta A}{\delta \ket{\Psi}} = 0 = \frac{\delta A}{\delta \langle\widetilde{\Psi}|},  \label{stationP}
\end{equation}
for all independent variations in the bra and ket states such that
$\delta\ket{\Psi(t_j)} = 0 =
\delta\langle\widetilde{\Psi}(t_j)|~;~j=0,1$, reproduces the
time-dependent Schr\"{o}dinger equations,
\begin{eqnarray}
H\ket{\Psi}=i\frac{\partial }{\partial t} \ket{\Psi}, ~~
\langle\widetilde{\Psi}|H = - i\frac{\partial }{\partial t}
\langle\widetilde{\Psi}|.        \label{schrod_eqns}
\end{eqnarray}
Equation (\ref{stationP}) gives Hamilton's equations of motion for
the cluster coefficients,
\begin{eqnarray}
i\frac{ds_I}{dt} = \frac{1}{N_I} \frac {\partial \langle H \rangle}{\partial
\widetilde{s}_I} ;~~ -i\frac{d\widetilde{s}_I}{dt} =
\frac{1}{N_I}\frac {\partial \langle H \rangle}{\partial s_I}, \label{Sevol}
\end{eqnarray}
where $N_I$ is the norm from Eq. (\ref{Cnorm}).

Following Ref.\cite{bi:cc} we make the following choice of
reference state and operators,
\begin{eqnarray}
\ket{\Phi_0} = \ket{0, \downarrow} ,
~~
\hat{S} = \hat{S_1} + \hat{S_2},                        \label{phiandS}
\end{eqnarray}
with
\begin{eqnarray}
\hat{S_1} = \sum _{n=1}^{\infty} s^{\left(1\right)}_n(t)
\left(b^\dagger \right)^n, ~~ \hat{S_2} = \sum _{n=1}^{\infty}
s^{\left(2\right)}_n(t) (b^{\dagger})^{n-1}\sigma^{+}.
\label{Sdefns}
\end{eqnarray}
The corresponding expansion for $\hat{\widetilde{S}}$ is
\begin{equation} \hat{\widetilde{S}} =
1+\hat{\widetilde{S}_1} + \hat{\widetilde{S}_2},
\end{equation}
with
\begin{eqnarray}
\hat{\widetilde{S}_1} = \sum _{n=1}^{\infty}
\widetilde{s}^{\left(1\right)}_n(t) b^n, ~~ \hat{\widetilde{S}_2}
= \sum _{n=1}^{\infty} \widetilde{s}^{\left(2\right)}_n(t)
b^{n-1}\sigma^{-}. \label{Stildefns}
\end{eqnarray}

We note that the states $\left(b^\dagger\right)^n|0,\downarrow\rangle$ and
$\left(b^\dagger\right)^{(n-1)}\sigma^+|0,\downarrow\rangle$ are eigenstates
of the excitation operator, $N$ with eigenvalues equal to $n$. They have
corresponding even or odd parity (i.e. eigenvalues of $\Pi$ equal to $+1$ or
$-1$, respectively) depending on whether $n$ is even or odd. The reference
state $\ket{\Phi_0}$ has positive parity.  This is the same as the
ground-state, as demanded by Eq.(\ref{PhiPsi}). The overlaps,
$N_I$, of the operators (\ref{Sdefns}) are
\begin{eqnarray} N^{(1)}_n =
\langle 0,\downarrow | b^n (b^\dagger)^n \ket{0,\downarrow}=n!~~~~~~~~~~~
\nonumber\\ N^{(2)}_n = \langle 0,\downarrow | \sigma^- b^{n-1}
(b^\dagger)^{n-1} \sigma^+\ket{0,\downarrow}=4(n-1)!~. \end{eqnarray}
This choice of reference state and cluster operators has been
found to predict a spurious phase transition at $g\rightarrow g_c\approx
0.665$ \cite{lo:li}.  This in fact signals a breakdown of the calculation
for $g>g_c$.  For $g<g_c$ this scheme gives excellent
results for the ground state, and so it is within this region that we shall
work.

\section{Time Evolution}

We shall start the system in the unexcited vacuum, $\ket{0,\downarrow}$.  This
corresponds to the initial, $t=t_0$, conditions
\begin{equation}
s^{(1)}_n(t_0) = s^{(2)}_n(t_0) = 0 
= \widetilde{s}^{(1)}_n(t_0) 
= \widetilde{s}^{(2)}_n(t_0).
\end{equation}
Other choices are certainly possible, but the above choice is the
most obvious one that corresponds to a physical state and which
satisfies Eq. (\ref{PhiPsi}).

\begin{figure}[tb]
\centerline{\includegraphics[clip=true,width=0.7\textwidth,angle=270]
            {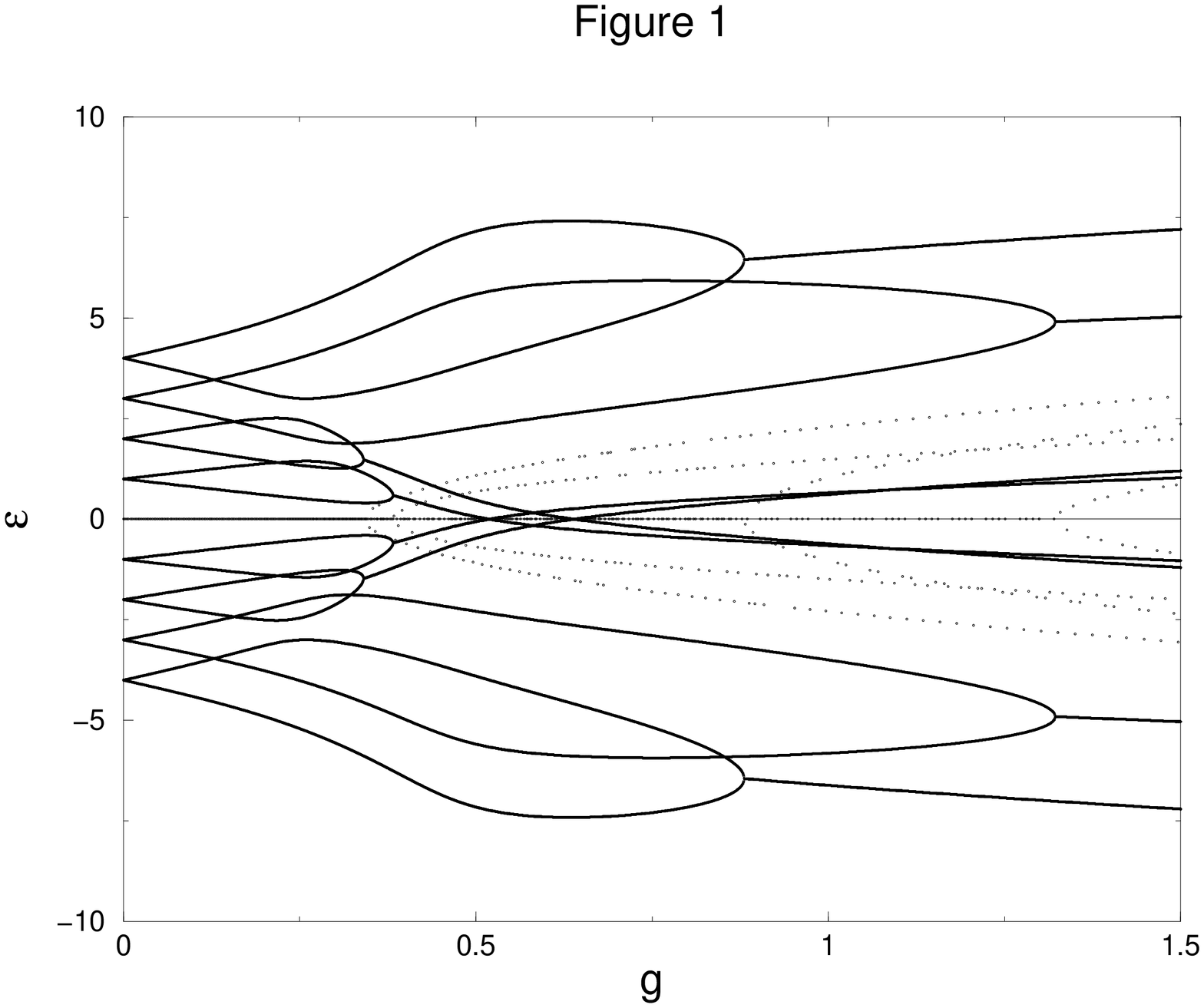}}
\caption{
The excitation-energy spectrum of the resonant Rabi Hamiltonian
($\omega=\omega_0=1$) as a function of $g$, as determined by a SUB-4 NCCM
calculation. Solid (dotted) lines show real (imaginary) parts of the
excitation energies. The negative-energy solutions and those with imaginary
components are spurious.
\label{F1}
 }
\end{figure}

In the evaluation of the time evolution equations we truncate the
sums in Eqs. (\ref{Sdefns}) and (\ref{Stildefns}) at $ n=N$, giving the
so-called SUB-$N$ approximation, in which all the coefficients $s^{(i)}_n(t)$
and $\widetilde{s}^{(i)}_n(t),~ i=1,2,$ for $n>N$ are set to zero.  The
resulting set of coupled equations (\ref{Sevol}) is then solved numerically by
taking finite time-steps.  We note that since the initial state
$|0,\downarrow\rangle$ has even parity $\Pi$, all the coefficients
$s^{(i)}_n(t)$ and $\widetilde{s}^{(i)}_n(t),~ i=1,2,$ vanish
identically for odd values of $n$, at all times.  The convergence of our
method with decreasing step size is demonstrated 
in Table \ref{table1}. Naively, one
would expect a similar type of convergence with increasing SUB-$N$ level.
Adding more cluster operators should allow the actual wavefunction to be
described more accurately. However, we find that this is only the case for
small coupling, as shown in Table \ref{table2}. In general 
we find that for a given
coupling parameter, $g$, there is a maximal value of the truncation index $N$
for which convergence occurs.  For higher values of $N$ the method diverges,
with the calculated cluster coefficients diverging to infinity.
To understand this behaviour we look at the energy spectrum
calculated by the NCCM under this scheme.  This spectrum is
determined by first calculating the ground-state coefficients,
$\left\{s_I,\widetilde{s_I}\right\}$, and using these to construct
the so-called dynamic matrix, $H_D$ of linear response theory \cite{bi:vl}.
This matrix is then diagonalised to give the energy eigenvalues. A full
discussion of this procedure would take us too far afield and so
we shall just consider the results and refer the reader to Ref.
\cite{bi:vl} for further information.  Fig. \ref{F1} shows the
excitation energies as determined by a SUB-4 calculation.  A
SUB-$N$ calculation yields 4$N$ eigenvalues, half of which are
spurious negative energy solutions, generated as a result of the
symmetry of $H_D$.

\begin{table}
\begin{center}
\begin{tabular}{|c|c|c|c|c|}
\hline Step size & $\Re\left\{s^{(2)}_2\right\}$ &
$\Im\left\{s^{(2)}_2\right\}$ & $\Re\left\{s^{(2)}_8\right\}$ &
$\Im\left\{s^{(2)}_8\right\}$ \\ \hline \hline
0.0333 & -9.168663$\times 10^{-3}$ & -1.772499$\times 10^{-2}$ & 2.946230$\times 10^{-8}$ & -2.517358$\times 10^{-8}$ \\
0.0250 & -9.167811$\times 10^{-3}$ & -1.772437$\times 10^{-2}$ & 2.365733$\times 10^{-8}$ & -2.493248$\times 10^{-8}$ \\
0.0200 & -9.167578$\times 10^{-3}$ & -1.772420$\times 10^{-2}$ & 2.212346$\times 10^{-8}$ & -2.482074$\times 10^{-8}$ \\
0.0100 & -9.167426$\times 10^{-3}$ & -1.772409$\times 10^{-2}$ & 2.114473$\times 10^{-8}$ & -2.473925$\times 10^{-8}$ \\
0.0050 & -9.167416$\times 10^{-3}$ & -1.772408$\times 10^{-2}$ & 2.108448$\times 10^{-8}$ & -2.473400$\times 10^{-8}$ \\
0.0033 & -9.167416$\times 10^{-3}$ & -1.772408$\times 10^{-2}$ & 2.108127$\times 10^{-8}$ & -2.473372$\times 10^{-8}$ \\
0.0025 & -9.167416$\times 10^{-3}$ & -1.772408$\times 10^{-2}$ & 2.108073$\times 10^{-8}$ & -2.473367$\times 10^{-8}$ \\
0.0010 & -9.167416$\times 10^{-3}$ & -1.772408$\times 10^{-2}$ & 2.108048$\times 10^{-8}$ & -2.473365$\times 10^{-8}$ \\
0.0005 & -9.167416$\times 10^{-3}$ & -1.772408$\times 10^{-2}$ & 2.108048$\times 10^{-8}$ & -2.473365$\times 10^{-8}$ \\
\hline
\end{tabular}
\end{center}
\caption{
Convergence of the method with decreasing step-size.  The
values of the real and imaginary parts of two typical coefficients
are shown at time $gt=1$, for a SUB-12 calculation with $g=0.05$.
\label{table1}
}
\end{table}

\begin{table}
\begin{center}
\begin{tabular}{|c|c|c|c|c|}
\hline
$N$
& $\Re\left\{s^{(2)}_2\right\}$
& $\Im\left\{s^{(2)}_2\right\}$
& $\Re\left\{s^{(2)}_8\right\}$
& $\Im\left\{s^{(2)}_8\right\}$ \\
\hline \hline
2  & -9.050635$\times 10^{-3}$ & -1.783754$\times 10^{-2}$ & - & - \\
4  & -9.167532$\times 10^{-3}$ & -1.772413$\times 10^{-2}$ & - & - \\
6  & -9.167416$\times 10^{-3}$ & -1.772408$\times 10^{-2}$ & - & - \\
8  & -9.167416$\times 10^{-3}$ & -1.772408$\times 10^{-2}$ & 1.675505$\times 10^{-8}$ &  4.392506$\times 10^{-9}$\\
10 & -9.167416$\times 10^{-3}$ & -1.772408$\times 10^{-2}$ & 2.126271$\times 10^{-8}$ & -2.456155$\times 10^{-8}$ \\
12 & -9.167416$\times 10^{-3}$ & -1.772408$\times 10^{-2}$ & 2.108048$\times 10^{-8}$ & -2.473365$\times 10^{-8}$ \\
14 & -9.167416$\times 10^{-3}$ & -1.772408$\times 10^{-2}$ & 2.107987$\times 10^{-8}$ & -2.473357$\times 10^{-8}$ \\
16 & -9.167416$\times 10^{-3}$ & -1.772408$\times 10^{-2}$ & 2.107987$\times 10^{-8}$ & -2.473356$\times 10^{-8}$ \\
18 & -9.167416$\times 10^{-3}$ & -1.772408$\times 10^{-2}$ & 2.107987$\times 10^{-8}$ & -2.473356$\times 10^{-8}$ \\
20 & -9.167416$\times 10^{-3}$ & -1.772408$\times 10^{-2}$ & 2.107987$\times 10^{-8}$ & -2.473356$\times 10^{-8}$ \\
40 & -9.167416$\times 10^{-3}$ & -1.772408$\times 10^{-2}$ & 2.107987$\times 10^{-8}$ & -2.473356$\times 10^{-8}$ \\
\hline
\end{tabular}
\end{center}
\caption{
Convergence of the SUB-$N$ method with increasing
truncation index $N$.  The values of the real and imaginary parts
of two typical coefficients are shown at time $gt=1$, for
$g=0.05$. The step size was 0.0005. This convergence is only valid
for small couplings.
\label{table2}
}
\end{table}

\begin{figure}[tb]
\centerline{\includegraphics[clip=true,width=0.7\textwidth,angle=270]
            {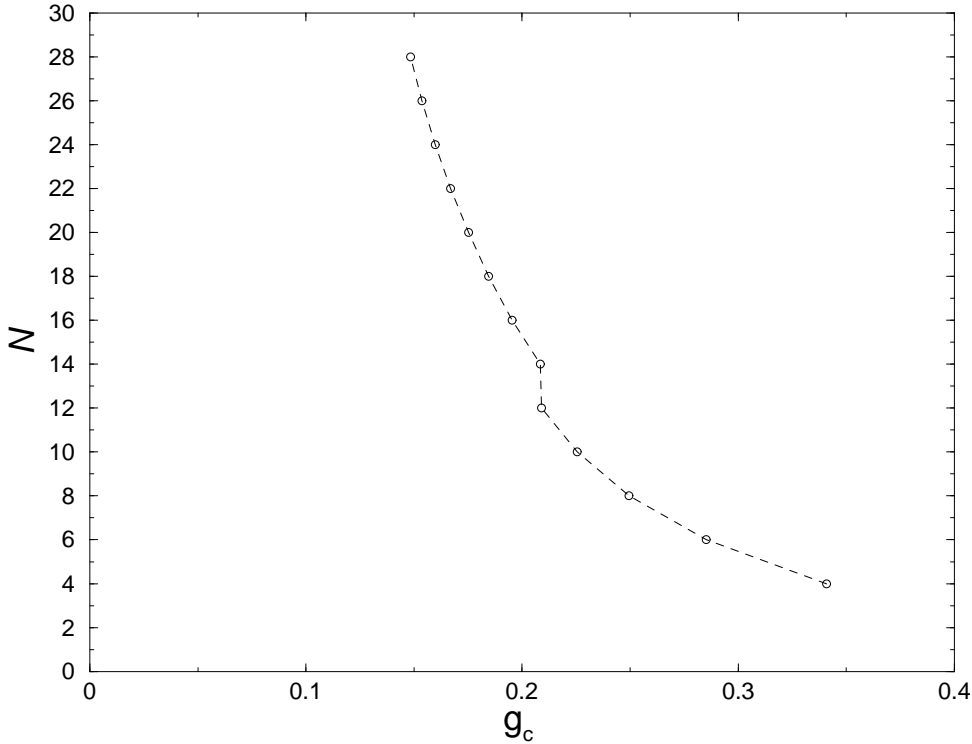}}
\caption{
The critical value $g_c=g_c(N)$ of the coupling parameter, above
which the NCCM time-evolution calculations break down, is shown
for various values of the SUB-$N$ truncation index $N$.  This is
the coupling above which the SUB-$N$ excitation spectrum develops imaginary
components.  The Hamiltonian is resonant ($\omega=\omega_0=1$),
\label{F2}
}
\end{figure}

The important feature to note is that at certain values of $g$
(e.g., at $g\approx 0.35$ for the SUB-4 case in Fig. \ref{F1}), 
which depend on the
truncation index $N$, two real energy levels come together, and as $g$
increases, become a complex conjugate pair.  Thus, for higher values of $g$,
the NCCM predicts a pair of complex conjugate energy eigenvalues for the
Hamiltonian. We can see what this means for the time evolution by considering
the basic quantum mechanical expression for the evolution of an arbitrary
wavefunction,
\begin{equation}
\ket{\Psi (t)} = \sum_n \langle u_n \ket{\Psi(t_0)}
e^{-iE_n(t-t_0)}\ket{u_n},        \label{QMevol}
\end{equation}
where $\ket{u_n}$ is the eigen-ket corresponding to energy
eigenvalue $E_n$. We see that having imaginary components to $E_n$
of both signs will lead to exponentially growing terms in this
sum, as opposed to bounded oscillations. Thus we expect our NCCM
time evolution calculations to break down for a given coupling if
the SUB-$N$ level predicts complex energy eigenvalues at that coupling.
Fig. \ref{F2} shows the values of $g$, for a given SUB-$N$ truncation
index $N$, above which the spectrum contains imaginary components.
These are the points where our time-evolution calculations break
down. The SUB-2 time evolution breaks down at exactly the point
where the ground-state calculation does.

Sree Latha and Durga Prasad have encountered similar problems in their 
applications of the closely related multi-reference time-dependent 
coupled cluster method (MRTDCCM) \cite{sl:p2}, which also posits an 
exponential Ansatz for the wavefunction, but uses a model space spanned by a 
number of states as opposed to the NCCMs single reference state. 
They trace the origin of the complex eigenvalues back to the use of the 
similarity transformed Hamiltonian $\tilde{H}$, which, when truncated, 
is not necessarily Hermitian.  This lack of Hermiticity permits complex
eigenvalues, whose corresponding eigenvectors are related to the the so-called
 ``intruder states''.  They conjecture that this situation may be 
expected to arise when the reference state or space interacts strongly 
with a state or set of states in the rest of the Hilbert space; which is 
just the situation  we observe here. 

\begin{figure}[tb]
\centerline{\includegraphics[clip=true,width=0.7\textwidth,angle=270]
            {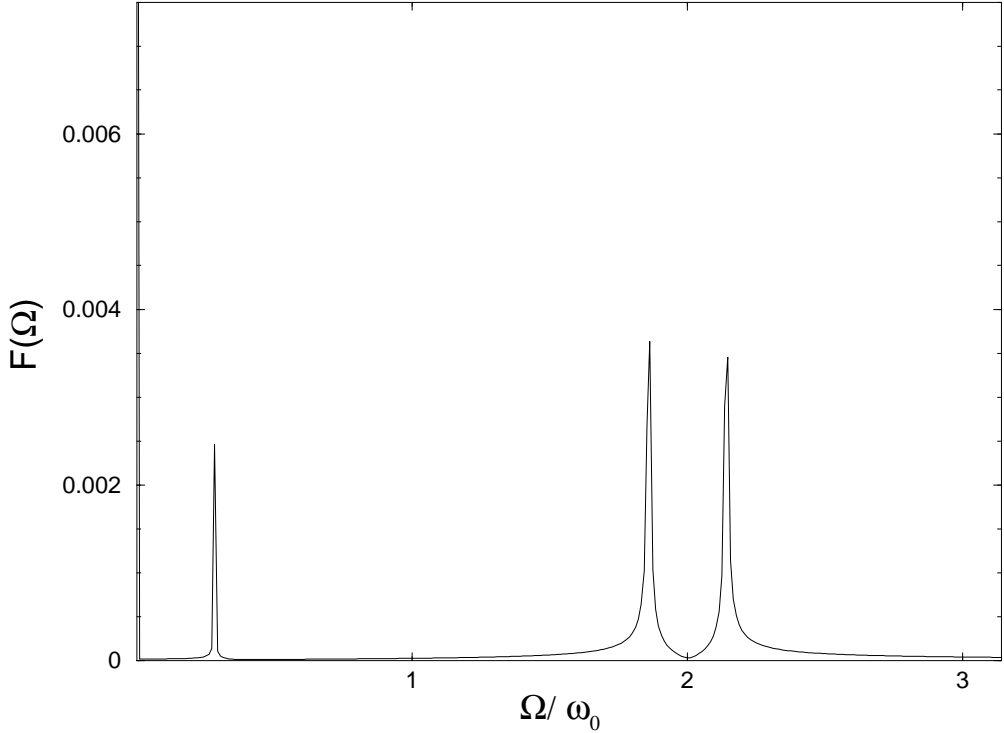}}
\caption{
Fourier transform of the quantity $f(t)$ from Eq.
(\ref{f(t)}) for the resonant Hamiltonian ($\omega=\omega_0=1$) with $g=0.05$.
This shows a discrete spectrum with three main frequency components.
\label{F3}
 }
\end{figure}

\begin{figure}[tb]
\centerline{\includegraphics[clip=true,width=0.7\textwidth,angle=270]
            {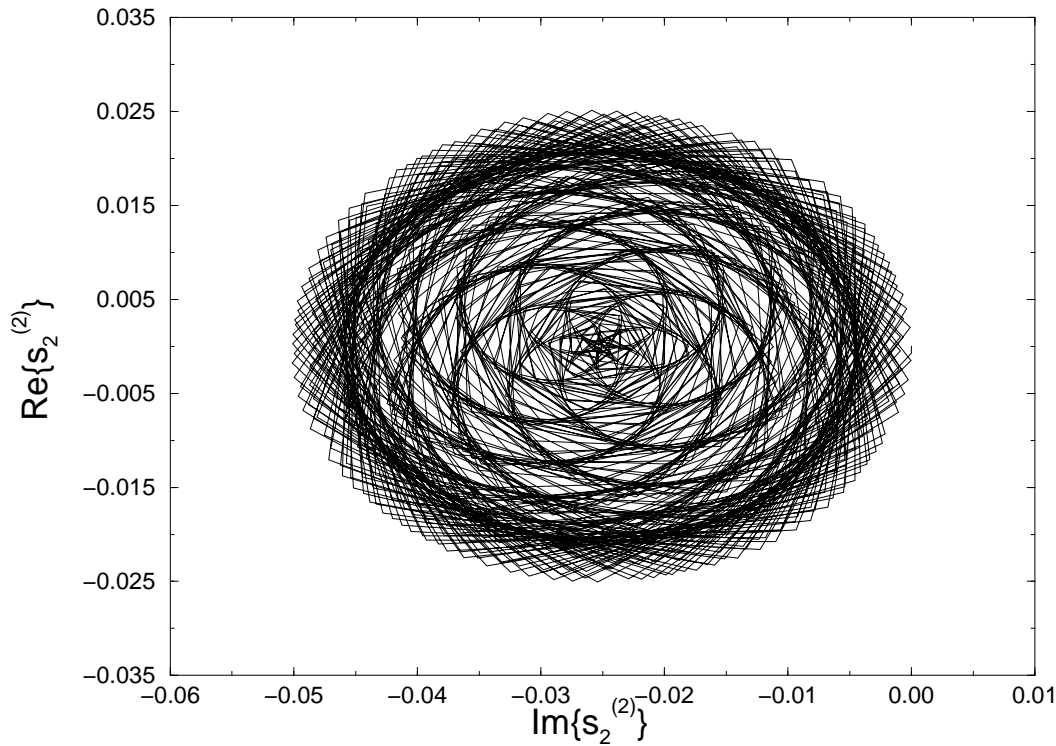}}
\caption{
A parametric ``phase-space'' plot of the coefficient
$s_2^{(2)}(t)$, shown for a range of the time parameter $t$ for
which $0\le gt \le 120$. ($g=0.05$, SUB-30, $\omega=\omega_0=1$)
\label{F4}
 }
\end{figure}

In understanding the following results, it will be useful to study
a Fourier transforms of the time series. Terms similar to
\begin{equation} f(t) = \sum_{k=1}^{N} (k-1)!~s^{(2)}_k(t)
\widetilde{s}^{(2)}_k(t)                        \label{f(t)}
\end{equation}
occur in all the quantities that we look at below. We define
$F(\Omega)$ as the Fourier transform of $f(t)$ and Fig. \ref{F3} shows
a plot of this quantity for typical parameters. We see a discrete
spectrum with three main peaks. These peaks correspond to the
lowest three positive-parity energy levels in the spectrum, as one
would expect.  This means that the time evolution will be
quasi-periodic. This quasi-periodic structure is reflected in the
behaviour of the cluster coefficients, as can be clearly seen from
the typical parametric plot shown in Fig. \ref{F4}.

\section{Atomic Inversion}
\begin{figure}[tb]
\centerline{\includegraphics[clip=true,width=0.7\textwidth]
            {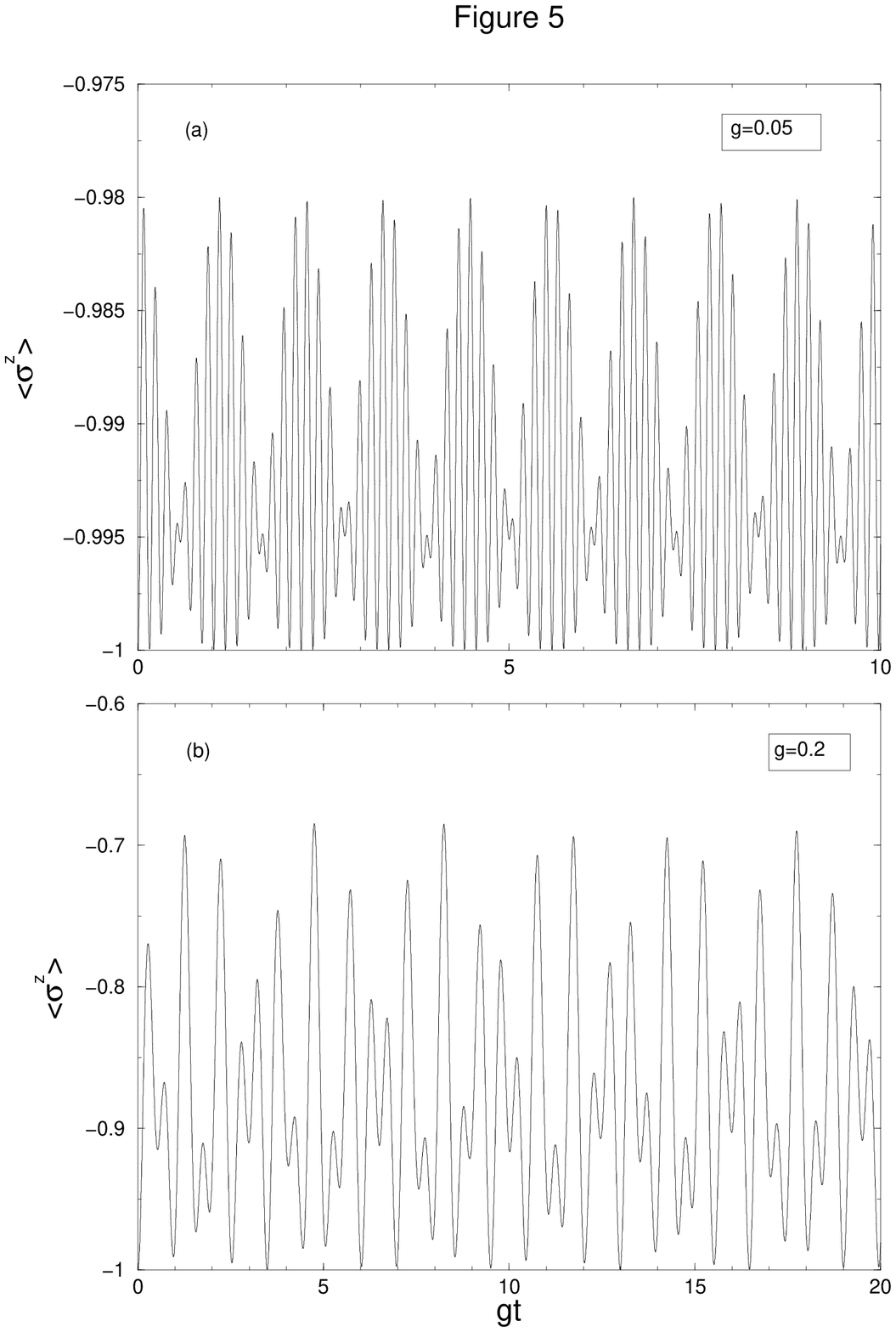}}
\caption{
Time evolution of the atomic inversion, $\langle \sigma^z
\rangle$, for two different couplings: a) $g=0.05$, SUB-30 and b)
$g=0.2$, SUB-14. The Hamiltonian is resonant, ($\omega = \omega_0=1$).
\label{F5}
 }
\end{figure}

\normalsize
The atomic inversion, $\langle\sigma^z\rangle$, has been the
primary atomic quantity of interest when studying the Rabi
system, not least because it is experimentally determinable
\cite{re:wa}.  In our NCCM scheme the atomic inversion is given
by
\begin{equation}
\langle\sigma^z\rangle = -1+8\sum_{n=1}^{\infty} (n-1)!~s^{(2)}_n
\widetilde{s}^{(2)}_n.
\label{AI} \end{equation}
Being an observable, $\langle\sigma^z\rangle$ should always be
real but the cluster-coefficients $s^{(i)}_n$ and $\widetilde{
s}^{(i)}_n$, $i=1,2$, are, in general, complex. The truncation of
the cluster operators leads to the exact hermiticity of the bra
and ket states being broken. This in turn means that the atomic
inversion calculated under the NCCM is not constrained to be real.
Calculations show that $s^{(2)}_2\widetilde{s}^{(2)}_2$ is the
dominant contribution to the sum in Eq. (\ref{AI}) and reveal
these two coefficients to be almost complex conjugate to one
other. Subsequent terms in the summation conspire to reduce the
size of the spurious imaginary part, this reduction becoming more
perfect with increasing truncation.  For small couplings we can
almost completely eliminate this imaginary part, although the
restriction on the maximum SUB-$N$ level permissible for higher
couplings means that we cannot eliminate this component entirely
as $g$ becomes larger. The complex part of the calculated
$\langle\sigma^z\rangle$ is small and can be used as a rough
measure of the error in the NCCM calculations. For the parameters
used here,
$\mathrm{RMS}\left(\Im\left\{\langle\sigma^z\rangle\right\}\right)
\approx 2\times10^{-13}$, 
$\mathrm{Max}\left(\Im\left\{\langle\sigma^z\rangle\right\}\right)
\approx 2 \times 10^{-11}$ 
for $g=0.05$ and,
$\mathrm{RMS}\left(\Im\left\{\langle\sigma^z\rangle\right\}\right)
\approx 1 \times10^{-6}$, 
$\mathrm{Max}\left(\Im\left\{\langle\sigma^z\rangle\right\}\right)
\approx 4 \times 10^{-4}$ for 
$g=0.2$. 
Figure \ref{F5} shows the evolution of the atomic inversion for two
different couplings, $g=0.05$ and $g=0.2$.
\section{Field Observables}
\subsection{Photon Number}
\begin{figure}[tb]
\centerline{\includegraphics[clip=true,width=0.7\textwidth]
            {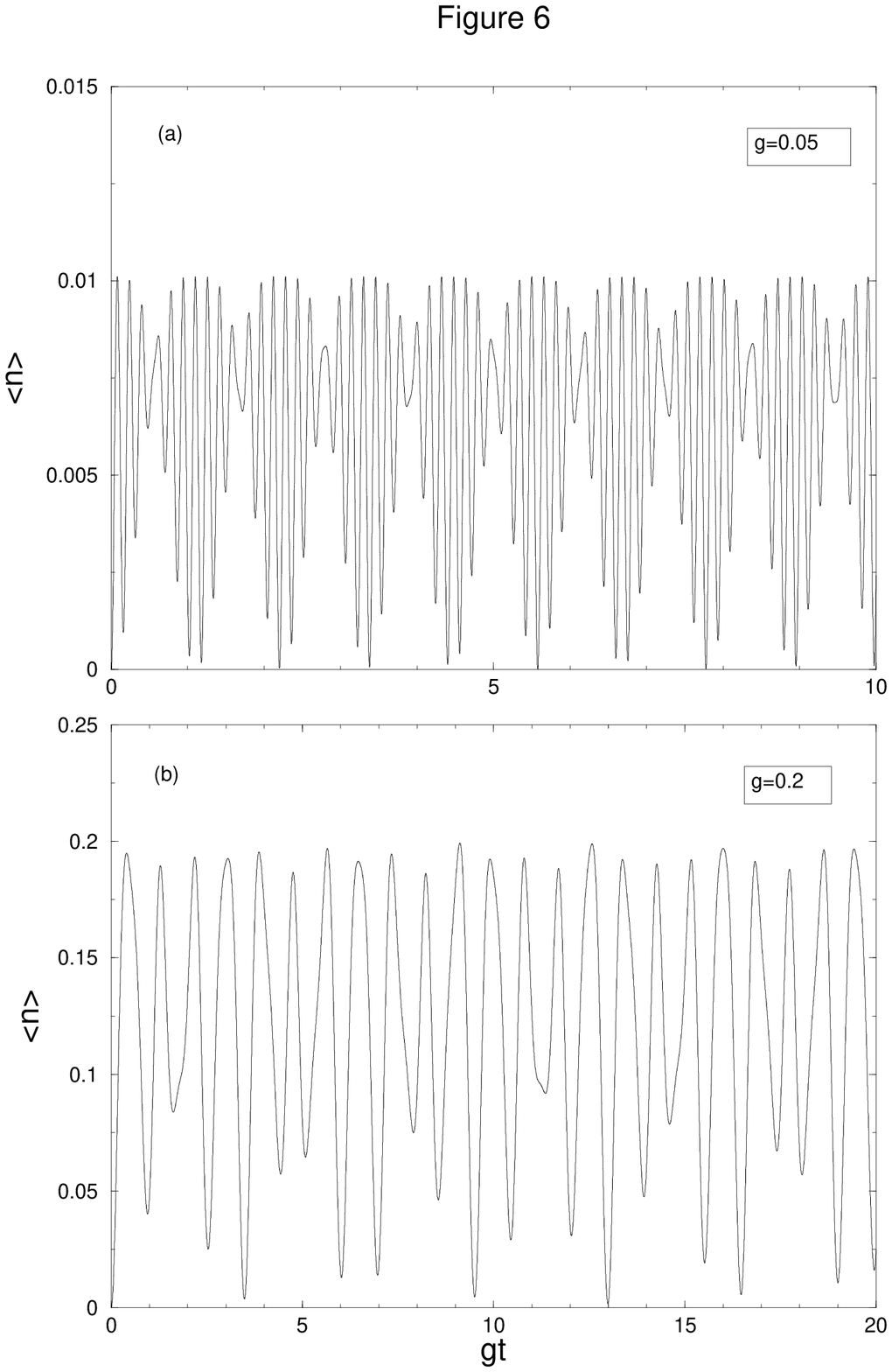}}
\caption{
Time evolution of the field occupation number $\langle n \rangle
\equiv \langle b^{\dagger} b \rangle$, using the same parameters
as in Fig. 5.  The Hamiltonian is resonant, ($\omega = \omega_0=1$).
\label{F6}
 }
\end{figure}

The most important operator associated with the field is
$b^{\dagger}b$, the photon number operator.  In terms of the NCCM
coefficients, it has an expectation value $\bar{n}(t)$,
\begin{equation}
\bar{n}(t)=\langle b^{\dagger}b \rangle = \sum_{n=1}^\infty n~n!~
s^{(1)}_n \widetilde{s}^{(1)}_{n} + 4\sum_{n=1}^\infty
(n-1)(n-1)!~ s^{(2)}_n \widetilde{s}^{(2)}_{n}. \label{bdagb}
\end{equation}
Figure \ref{F6} shows the evolution of this quantity.  The nature of this
evolution is very similar to that of the atomic inversion, and we clearly
observe the atom exchanging energy with the field.  A time average of $\langle
b^{\dagger}b \rangle$ provides an estimate of the mean number of virtual
photons in the cavity at any given moment.  For example, for $g=0.05$,
$\overline{ \langle b^{\dagger}b \rangle}\approx 0.0063$ and for $g=0.2$,
$\overline{ \langle b^{\dagger}b \rangle}\approx 0.12$.
\subsection{Photon Anti-bunching}
The next field observable we shall study is the second-order
correlation function,
\begin{equation}
g^{(2)}(\tau) = \frac{\langle b^{\dagger}(t)
b^{\dagger}(t+\tau)b(t+\tau)b(t)\rangle}{\langle
b^{\dagger}(t)b(t)\rangle^2}. \label{g2tau}
\end{equation}
This allows us to study whether the field exhibits anti-bunching.
For this purpose, we only require $g^{(2)}(0)$,
\begin{equation}
g^{(2)}(0) = \frac{\langle b^{\dagger}
b^{\dagger}bb\rangle}{\langle b^{\dagger}b\rangle^2}, \label{g20}
\end{equation}
and we define the convenient parameter \cite{ru:go,ca:wa}
\begin{equation}
y\equiv\langle b^{\dagger} b^{\dagger}bb\rangle - \langle
b^{\dagger}b\rangle^2.                              \label{y}
\end{equation}
A value of $y<0$ corresponds to $g^{(2)}(0)<1$ and indicates an
anti-bunched field.

\begin{figure}[tb]
\centerline{\includegraphics[clip=true,width=0.7\textwidth]
            {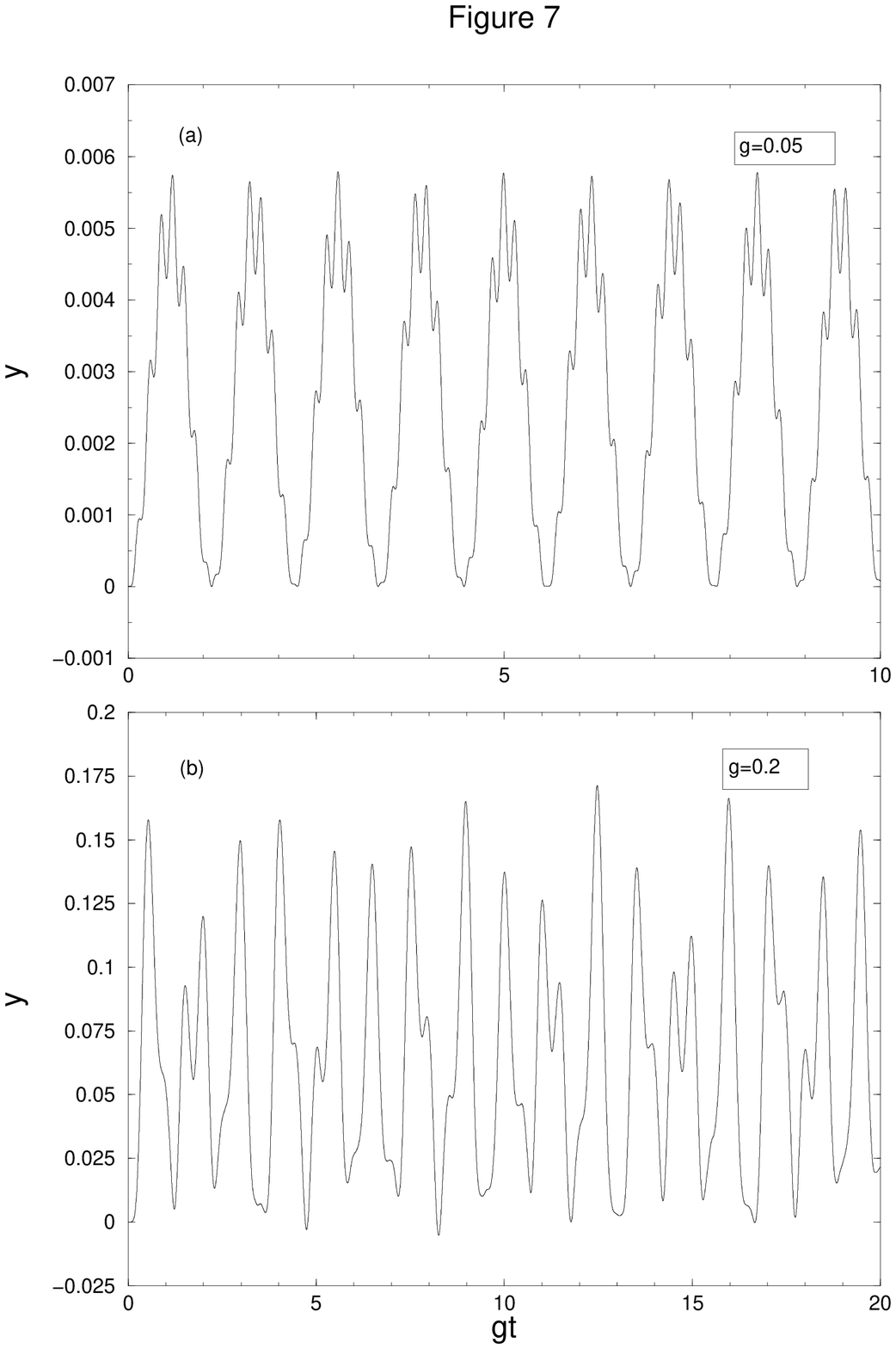}}
\caption{
Time evolution of the photon anti-bunching measure, $y$,
using the same parameters as Fig. 5.  Note that $y<0$ indicates
anti-bunching.  The Hamiltonian is resonant, ($\omega = \omega_0=1$).
\label{F7}
 }
\end{figure}

For the NCCM, $\langle b^{\dagger} b^{\dagger}bb\rangle$ is given
by
\begin{eqnarray}
\langle b^{\dagger} b^{\dagger}bb\rangle &=& \sum_{n=1}^\infty
n(n-1)n!~ s^{(1)}_n \widetilde{s}^{(1)}_{n}\nonumber \\
&+&\sum_{n,m=1}^\infty nm(n+m)!~ s^{(1)}_n s^{(1)}_m
\widetilde{s}^{(1)}_{n+m} \nonumber \\ &+& 4\sum_{n=1}^\infty
(n-1)(n-2)(n-1)!~s^{(2)}_n \widetilde{s}^{(2)}_{n} \nonumber \\
&+& 8\sum_{n,m=1}^\infty n(m-1)(n+m-1)!~ s^{(1)}_n
s^{(2)}_m\widetilde{s}^{(2)}_{n+m}.  \label{NCCMbdbdbb}
\end{eqnarray}
Figure \ref{F7} shows the quantity $y$ for various couplings.  For small
values of $g$ we see that $y$ never drops below zero and thus the
field is never anti-bunched.  For higher values of coupling, we
see that $y$ occasionally does fall very slightly below zero. This
effect is extremely small and we believe that it is due to
inaccuracies introduced by using small truncation levels.
\subsection{Squeezing}
\begin{figure}[tb]
\centerline{\includegraphics[clip=true,width=0.5\textwidth]
            {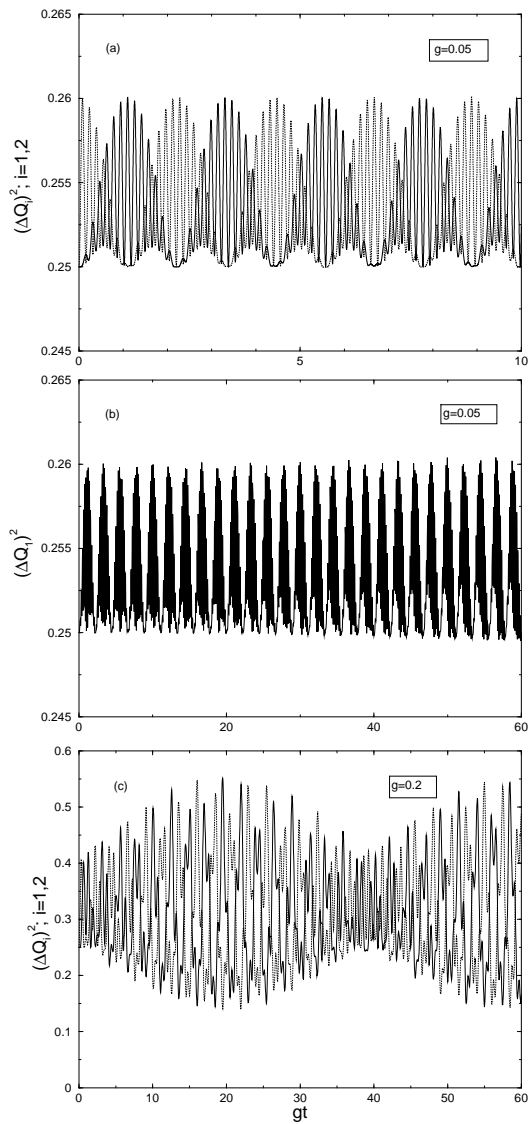}}
\caption{
Time evolution of the squeezing variances.
Squeezing occurs when $\left(\Delta Q_1\right)^2$ or $\left(\Delta
Q_2 \right)^2 < \frac{1}{4}$.  a) $g=0.05$, SUB-30, short time, b)
$g=0.05$, SUB-30, longer time, and c) $g=0.2$, SUB-14. The
Hamiltonian is resonant ($\omega = \omega_0=1$).
\label{F8}
 }
\end{figure}

By analogy to the position and momentum operators of the harmonic
oscillator, we introduce the following quadrature operators for
the electromagnetic field \cite{me:zu},
\begin{eqnarray}
Q_1&\equiv&  \frac{1}{2}\left[ b\exp{(i\omega
t)}+b^{\dagger}\exp{(-i\omega t)} \right],\nonumber\\ Q_2&\equiv&
\frac{1}{2i}\left[ b\exp{(i\omega t)}-b^{\dagger}\exp{(-i\omega t)}
\right]. \label{quads}
\end{eqnarray}
Their variances, given by
\begin{eqnarray}
\left( \Delta Q_1\right)^2=\frac{1}{2}\left\{\langle b^{\dagger} b
\rangle + \Re\left[ \langle b^2 \rangle\exp{(2i\omega
t)}\right]\right\} - \left\{\Re\left[ \langle b \rangle \exp{(i
\omega t)} \right] \right\}^2 + \frac{1}{4},\nonumber\\ \left(
\Delta Q_2\right)^2=\frac{1}{2}\left\{\langle b^{\dagger} b
\rangle - \Re\left[ \langle b^2 \rangle\exp{(2i\omega
t)}\right]\right\} - \left\{\Im\left[ \langle b \rangle \exp{(i
\omega t)} \right] \right\}^2 + \frac{1}{4}, \label{Qvars}
\end{eqnarray}
satisfy the uncertainty relation,
\begin{equation}
  \Delta Q_1 \Delta Q_2 \geq \frac{1}{4}.                     
  \label{uncert}
\end{equation}
The field is said to be squeezed when either $\left(\Delta
Q_1\right)^2$ or $\left(\Delta Q_2 \right)^2 < \frac{1}{4}$.  The
NCCM expressions for the remaining expectation values in Eq.
(\ref{Qvars}) are easily calculated, and are given by
\begin{eqnarray}
\langle b^2 \rangle &=& 
  2s^{(1)}_2+ s^{(1)}_1s^{(1)}_1
  + \sum_{n=3}^\infty n!~ s^{(1)}_n \widetilde{s}^{(1)}_{n-2}
    \nonumber \\
 &+&\sum_{n=1}^\infty\sum_{m=1}^\infty nm(n+m-2)!~ s^{(1)}_n s^{(1)}_m
    \widetilde{s}^{(1)}_{n+m-2} \rb{1 - \delta_{n,1}\delta_{m,1}}
    \nonumber \\ 
 &+& 4\sum_{n=3}^\infty (n-1)!~s^{(2)}_n 
    \widetilde{s}^{(2)}_{n-2} 
    \nonumber\\ 
 &+& 8\sum_{n=1}^\infty \sum^\infty_{m=2} n(m-1)(n+m-3)!~ s^{(1)}_n
     s^{(2)}_m\widetilde{s}^{(2)}_{n+m-2}, \\ 
\langle b \rangle &=&
  s^{(1)}_1 + \sum_{n=2}^\infty n!~ s^{(1)}_n
  \widetilde{s}^{(1)}_{n-1}
  + 4\sum_{n=2}^\infty (n-1)!~ s^{(2)}_n \widetilde{s}^{(2)}_{n-1}.     
\label{NCCMbs}
\end{eqnarray}
Figure \ref{F8} shows the evolution of these variances
from the vacuum.  For $g=0.05$, we see that the values of $(\Delta Q_i)^2;
~i=1,2$ only just drop below $\frac {1}{4}$; (min$(\Delta
Q_i)^2 = 0.2495$ for $g=0.05$). However, for greater couplings, we do
see that squeezing is more significant; (min$(\Delta Q_i)^2 = 0.1351$ for
$g=0.2$).

\section{Discussion}

We have demonstrated the ability of the NCCM to describe the time
evolution of a simple but important quantum system.  We have also
outlined the limitations of the method.  Despite which, we have 
been able to obtain a range of useful reults for the system.

We have seen that the counter-rotating terms give rise to quite
complex behaviour in the evolution from the unexcited vacuum. The
field, although not anti-bunched, does exhibit squeezing, which
becomes more pronounced with increased coupling. However, the
small absolute magnitude of these effects and the limitations of
the model, such as neglect of thermal photons, clouds the
experimental significance of the results.

In assessing the performance of the NCCM in describing this system, 
it is useful to compare results with those obtained by the 
configuration interaction (CI) method using the same basis.  The CI
 method is equivalent to diagonalisation in a truncated set of basis
states.  If we use the same number of states in both CI and NCCM
calculations, the two procedures are of approximately the same
computational complexity.  In Table \ref{table3} we compare atomic inversions
of the Rabi system calculated by both methods for two couplings.  We
have chosen times for this comparison where the atomic inversion is
near a maximum, and thus the state system at these times is as far from 
the reference state as possible.  It should be noted that both sets of 
$N=16$ results
from the CI method may be treated as exact to the precision 
of Table \ref{table3}, as they are
converged with results for much larger $N$.  These results demonstrate 
several things about the performance of the NCCM in this system.  In the
region where the NCCM spectrum contains no complex energies, the NCCM 
 describes the system better than the equivalent CI diagonalisation, 
especially for low ($N=2,4$) truncation levels.  This is due to the superior 
counting of independent excitations in the NCCM \cite{bi:vl}.  
Conversely, the $g=0.2$ results reflect 
the fact that the presence of the complex energies prevents the NCCM from 
converging, limiting the accuracy of the NCCM for higher couplings.

\begin{table}
\begin{center}
\begin{tabular}{|c|c|c|c|c|}
\hline
& \multicolumn{2}{|c|}{$g=0.05$; $gt =1.25$ } 
  & \multicolumn{2}{|c|}{$g=0.20$; $gt=1.26$}\\
  \cline{2-5}
N & CI & NCCM & CI & NCCM \\
\hline \hline
2  & -0.999999 & -0.981944 & -0.936100 & -0.745812 \\
4  & -0.981757 & -0.981759 & -0.696200 & -0.696457 \\
6  & -0.981759 & -0.981759 & -0.692675 & -0.694392 \\
8  & -0.981759 & -0.981759 & -0.693094 & -0.694211 \\
10 & -0.981759 & -0.981759 & -0.693087 & -0.692888 \\
12 & -0.981759 & -0.981759 & -0.693087 & -0.692909 \\ 
14 & -0.981759 & -0.981759 & -0.693087 & -0.692926 \\
16 & -0.981759 & -0.981759 & -0.693087 & -0.693384 \\
\hline
\end{tabular}
\end{center}
\caption{
 Comparison of NCCM SUB-$N$ results with results obtained from  CI
diagonalisation including the same number of basis states.  The table lists
the atomic inversion of the resonant Rabi Hamiltonian 
$(\omega = \omega_0 = 1)$ for  two different couplings.  The times, 
given by $gt$, where the inversion was evaluated, were chosen so that 
the inversion was near a local maximum. 
\label{table3}
}
\end{table}

The initial aim of applying the NCCM to the Rabi Hamiltonian was
to produce an accurate microscopic description of the time
evolution of the system across the whole coupling range.  This has
however not been completely realized due to the incursion of
complex energies in the NCCM spectrum. Similar problems occur in
the ground-state description. The underlying reasons for this
failure of the NCCM are not yet entirely clear to us, and will
merit further study.  However, it does seem likely that the
existence of the Juddian points, at which level crossings occur,
indicates the presence of a subtle symmetry whose preservation or
breaking is not reflected in our simple choice of reference state.
It should be noted that although van der Walt \cite{wa:ph} has
tried a number of different reference states and operator
selections, none of these has yet entirely solved this problem.

Future work includes analysis of the Rabi Hamiltonian with the
NCCM in the holomorphic representation, and an extension of the
above method to evolution from a coherent state.

\section{Acknowledgments}

C. E. acknowledges the financial support of a research studentship
from the Engineering and Physical Sciences Research Council (E.P.S.R.C.) of Great Britain.

\end{document}